# CMOS-compatible vanadium dioxide via Pulsed Laser and Atomic Layer deposition: towards ultra-thin film phase-change layers

*Anna Varini, Cyrille Masserey, Vanessa Conti, Zahra Saadat Somaehsofla, Ehsan Ansari, Igor Stolichnov, Adrian M. Ionescu\**

AUTHOR ADDRESS (Nanoelectronic Devices Laboratory, EPFL, 1015, Lausanne, Switzerland)

ABSTRACT

Vanadium dioxide (VO$_2$), a well-known Mott insulator, is a highly studied electronic material with promising applications in information processing and storage, including neuromorphic devices and circuits. While epitaxial VO$_2$ layers exhibit exceptional properties, such as a sharp and abrupt conductivity change at the metal-insulator transition, fabricating polycrystalline VO$_2$ films on silicon substrates often involves trade-offs in transport characteristics and switching performance, especially for ultra-thin layers required in advanced gate applications. In this study, we explore the growth of VO$_2$ films on standard CMOS-compatible wet-oxidized silicon wafers using two established deposition techniques: pulsed laser deposition (PLD) and atomic layer deposition (ALD). VO$_2$ films, ranging in thickness from 200 nm to less than 10 nm, were systematically characterized through structural and electrical analyses to optimize key growth parameters. Temperature and pressure were identified as the primary factors affecting VO$_2$ film quality, and the optimal growth conditions across the entire thickness range are discussed in detail. PLD and ALD each offer distinct advantages: PLD enables the formation of high-density films, while ALD allows for conformal deposition on complex 3D structures. We demonstrate that both methods can successfully produce ultra-thin VO$_2$ layers down to 6–8 nm with functional properties suitable for practical applications. This work underscores the potential of VO$_2$ for fully CMOS-compatible phase-change switching devices and provides valuable insights into optimizing growth processes for polycrystalline VO$_2$ films.

INTRODUCTION

Vanadium dioxide thin films demonstrate reversible change in electrical, optical and thermal properties[1–5] that makes them interesting candidates for integration in the wide range of optical and electronic devices, such as switches, modulators, memories, transistors[6–10] and in different types of sensors[11–14]. For example, by applying the appropriate strain, the lattice structure of the VO$_2$ material can be changed, resulting in a phase transition. Temperature control, optical control, electric field control, and strain control can all effectively use the semiconductor-to-metal phase transformation characteristics of VO$_2$. Reversible change of VO2 properties is based on phase transition mechanism that is currently explained by combination of Mott-Hubbard phase transition driven by strong electron correlation and the Peierls phase transition driven by the change of the lattice structure [15–18]. Experimental understanding of stimulus response of $VO_2$ film to light fields, electrostatic fields, terahertz pulses, or stresses may also complement the understanding of MIT mechanisms [19]. While there are multiple methods of growing $VO_2$ (Sol-gel, Hydrothermal, MBE, MOCVD, CVD, Sputtering, Magnetron, PLD, ALD) it is difficult to guaranty high-purity and high-quality $VO_2$ films, simply because V ion in $VO_2$ is not the highest valence state. Crystal structure and purity of $VO_2$ will vary depending on the preparation method, including $VO_2$ (R), $VO_2$ (M1), $VO_2$ (M2), $VO_2$ (A), $VO_2$ (B), $VO_2$ (C), $VO_2$ (D), $VO_2$ (T), etc.[20,21]. Moreover, for the integration in CMOS domain $VO_2$ thin film has to be grown on standard Si wafer with $SiO_2$ top layer that has intrinsic build in strain and thermal expansion coefficient mismatch with $VO_2$. Thin films grown on $Si/SiO_2$ substrates are polycrystalline and will demonstrate smaller switching ration compared to the epitaxial films grown on lattice matching substrates such as sapphire, $TiO_2$ or Mica[22–24]. Nevertheless, high switching ration is not always needed to enable application of $VO_2$ films for specific devices. There are multiple reports of lower switching ration films used in $VO_2$ sensors and other devices[25–27]. Rather than aiming for the highest on/off ratio it is more important to understand what performance can be expected from a given thickness of $VO_2$ film to be able to design future applications more accurately. It is known that $VO_2$ properties are strongly affected by the substrate temperature, reaction gas composition, gas flow rate, and other factors that have to be well control. In this work we use $Si/SiO_2$ CMOS compatible substrate and demonstrate how control of growth parameters in PLD and ALD recipes influence thin film properties. Interestingly, we have observed that IMT and MIT value, as well as temperature of hysteresis window, are shifting towards higher values in the films below 100nm. We attribute this shift to intrinsic strain of $SiO_2$ layer and thermal expansion coefficient mismatch of $SiO_2$ and $VO_2$. This result alone demonstrates doping like tuning of switching temperature. Gang Xu et al.[28] used radiofrequency reactive magnetron sputtering for epitaxial growth of $VO_2$ thin films on α-$Al_2O_3$ (0001) sapphire substrate and investigated the effect of 3–150 nm film thickness effect on optical properties. His results showed that as the film thickness decreased, the crystal metal phase transition temperature of $VO_2$ thin films significantly decreased. In our case of PLD and ALD non epitaxial $VO_2$ growth the shift in phase transition temperature is to the opposite direction and with both IMT and MIT increasing for thinner films. This difference can be explained by the fact that α-$Al_2O_3$ (0001) substrate has matching lattice structure and distortion will increase for thicker films, while in our case initial distortion will have less of an effect for thicker films. Jiang Meng et al.[29] has also found that by accurately adjusting the oxygen flow ratio without doping elements, the phase transition temperature can be controlled between 46°C and 72°C for $VO_2$ thin films synthesized on quartz glass. Therefore, our work contributes to the understanding of the phase transition temperature shift for the wide range of $VO_2$ thin films grown on CMOS compatible wafers. However, important highlight of our work is that we managed to achieve continues film below 10nm of thickness with both PLD and ALD method that has not been demonstrated yet according to our best knowledge.  There is lack of detailed research on the film formation mechanism for PLD and ALD methods, however the challenge of growing continues films below 10nm of thickness is well documented and normally attributed to the initial growth of island -like particles (formation and growth of crystal nuclei) and only consequent formation of a continuous film for the films of higher thickness[30]. We have partially overcome this limitation by using slow deposition rates and allowing longer time for the clusters to join in the continues film under carefully controlled annealing conditions.

DESIGN OF EXPERIMENTS AND RESULTS

Standard {100} oriented 525um Si wafers with 200nm WetOx ($SiO_2$) were used as substrates for all grown films. We report optimization strategy for both PLD and ALD growth methods that resulted in ultra-thin polycrystalline films with switching rations that is low but useable in $VO_2$ devices. Our approach was based on review of previously available data and narrowing down the variability range of key growth parameters. Each key parameter was varied in that narrow range while all other parameters were kept fixed. Such method might be long and expensive, but it is the best way to see non-linear dependances of multi-parameter process and real influence of each evaluated parameter. Bellow you can find step by step report on PLD and ALD experiments. Quality of resulting films was evaluated by means of four-point prob measurements inside temperature chamber, where temperature was cycled between 40°C and 90°C to determine switching on/off ratio and hysteresis of the film. XRD measurement were also used to confirm pure $VO_2$ phase or presence of a different $V_xO_y$ phases. Once best film was identified for a given variable parameter that parameter was fixed and investigation was moved into next influential parameter. The order of testing was decided based on previous results that indicated temperature as main factor of influence, followed by pressure and laser energy/frequency or fluence of deposition.

PULSED LASER DEPOSITION METHOD

Pulsed laser deposition (PLD) is recognized as a fast, clean, and versatile physical vapor deposition technique, capable of producing high-crystal-quality films. Its ability to precisely control key parameters—such as laser energy, pulse frequency, substrate temperature, deposition time, total chamber pressure, and partial gas pressure—makes it particularly suitable for growing complex oxides like vanadium dioxide ($VO_2$). The fine-tuning of these parameters allows for the careful modulation of film properties, including thickness uniformity, crystallinity, and stoichiometry, all of which are crucial for optimizing the metal-insulator transition (MIT) characteristics of $VO_2$ films.

PLD's unique advantage lies in its ability to maintain stoichiometric transfer from the target to the substrate, even for multi-component materials, due to the highly energetic nature of the laser-induced ablation plume. This feature ensures the faithful reproduction of the target's composition in the deposited film, a critical factor for $VO_2$, which is sensitive to deviations in vanadium and oxygen ratios. Furthermore, PLD is particularly effective in producing epitaxial and polycrystalline films with minimal contamination, making it an environmentally friendly option compared to chemical vapor deposition (CVD) or sputtering techniques, which may involve complex precursors or produce harmful byproducts.

Recent advances in PLD have demonstrated its capability to integrate functional dopants into $VO_2$, enabling the engineering of its phase-transition temperature and electronic properties. For instance, doping[31–35] with elements such as chromium (Cr), tungsten (W), niobium (Nb) and germanium (Ge) has been shown to modulate the MIT temperature, expanding the potential for $VO_2$ in temperature-tunable devices. This flexibility opens avenues for tailoring $VO_2$ for specific applications, such as tunable infrared optics, sensors, and neuromorphic computing elements. The straightforward incorporation of dopants via PLD, facilitated by the ability to alternate targets or introduce reactive gases during deposition, provides a promising route for fine-tuning $VO_2$'s electrical and optical properties. This capability, combined with the high control over deposition conditions, positions PLD as a state-of-the-art technique for advancing $VO_2$-based technologies.

TEMPERATURE EFFECT

First most important parameter that has influenced PLD growth was temperature during deposition and additional annealing. With critical minimum temperature required to form polycrystalline rather than amorphous layer during deposition. In our work continuous polycrystalline layers with high on/off ration started to form at 400°C . Increasing temperature of deposition have resulted in a different morphology with more granular and rough film structure with lower on/off ratio. Fig.1 shows SEM micrographs of samples (a) to (d) grown at fixed chamber pressure of 7.5 mTorr with 15 sccm of Oxygen, laser frequency of 28Hz, laser energy 230mJ and varied temperatures of 300°C, 400°C, 500°C  and 600°C  respectively. Every grown film was characterized by four-point prob measurement inside temperature chamber where temperature was cycled multiple times between 40°C and 90°C. Fig. 2. shows comparison of switching behavior for films grown udder identical conditions apart from chamber temperature and XRD results for three switching polycrystalline films. From the comparison of XRD and switching behavior, it can be seen that sample (b) grown at 400°C  has $VO_2$ phase as well as highest on/off ration and lowest hysteresis.

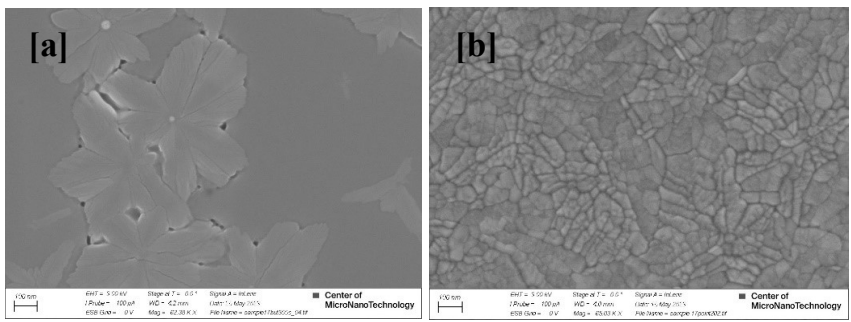

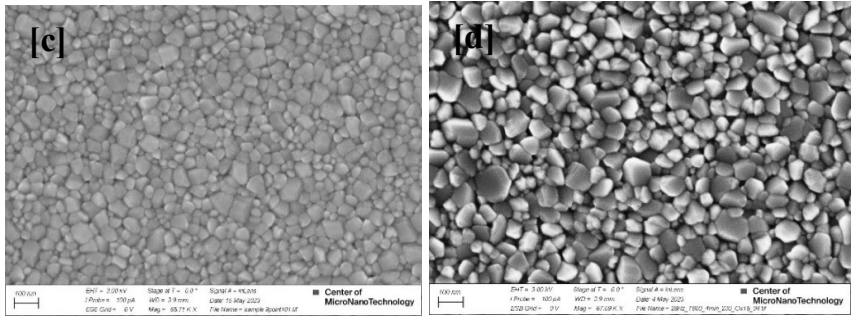

Fig. 1. SEM micrographs of VO$_2$ films grown at (a) 300°C (b) 400°C (c) 500°C (d) 600°C. Other growth parameters were fixed for all 4 samples: P=7.5 mTorr,15 sccm of Oxygen, laser frequency=28Hz, laser energy 230mJ, laser fluence=1.01mJ/cm2. Scale bar 100nm.

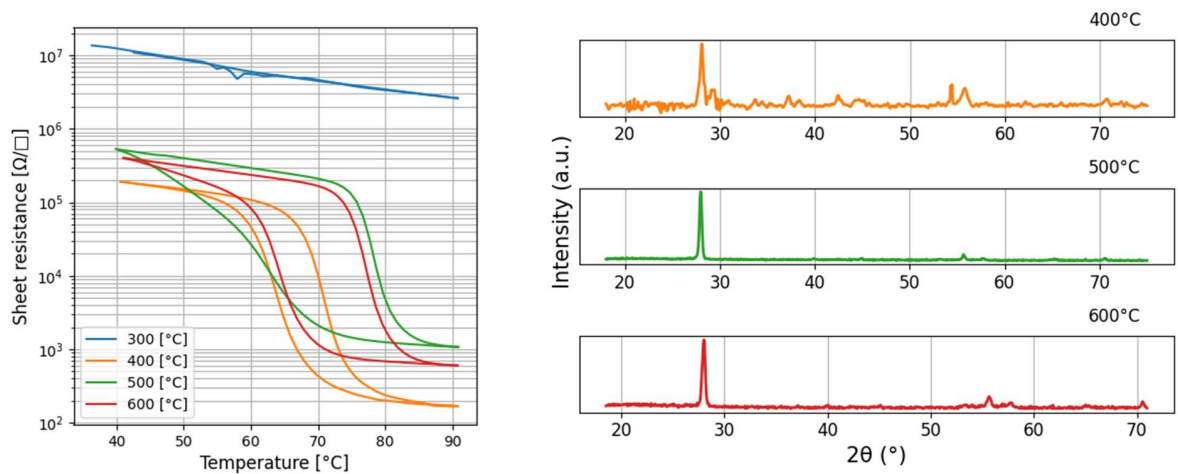

Fig.2. Left: Temperature sweep on samples (a) to (d) measured with pour-point prob configuration. Right: XRD spectrum for samples (b) to (d), sample (a) is not measured by XRD due to amorphous nature of the film.

Therefore, 400C was fixed as the best temperature and to further investigate influence of temperature sample (b) was annealed for 10min at 500C and 600C. SEM micrographs and respective before and after annealing switching curves for sample (b) can be seen on Fig.3.

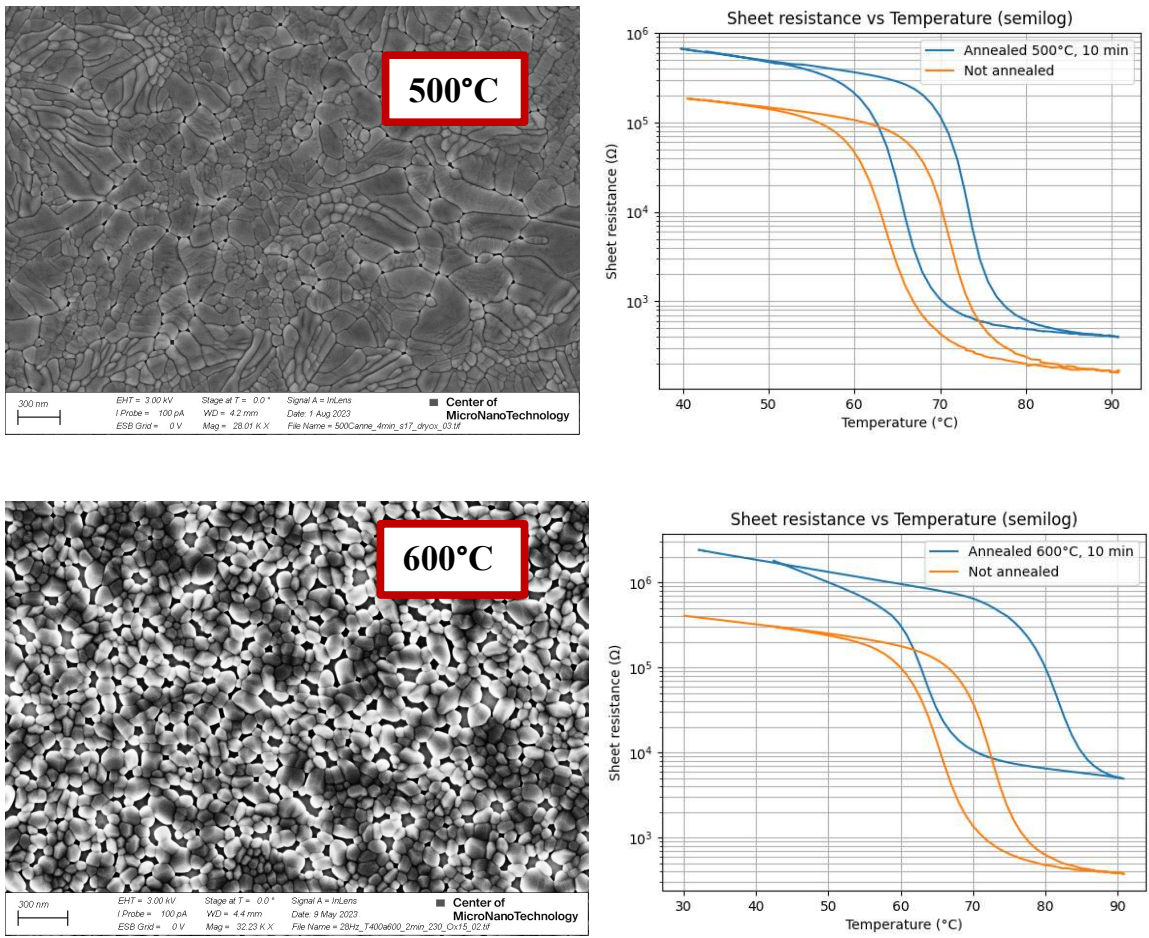

Fig.3. SEM micrographs and respective before and after annealing switching curves for sample (b). Top part of figure is for annealing at 500°C during 10min and bottom part of figure is for annealing at 600°C for 10min. Other growth parameters: P=7.5 mTorr, 15 sccm of Oxygen, laser frequency=28Hz, laser energy 230mJ, laser fluence=1.01 mJ/cm2. Scale bar 300nm

It can be seen from Fig.3 that annealing at 500°C has resulted in higher on/off ratio and smaller hysteresis of the film, while annealing at 600°C had a negative effect on the film properties resulting in bigger hysteresis and smaller on/off ratio. Other annealing temperatures were also tried and can be available in supplementary materials. The best results were achieved for 10 min of annealing at 500°C. Morphology of films annealed at higher temperatures resemble the films directly deposited at higher temperatures, but films grown directly at 500°C have lower on/off ratio compared to films annealed at 500°C.

PRESSURE EFFECT

It has been reported that oxygen partial pressure has an impact on the valence state of the final vanadium oxide film and on transition behavior[36], however, magnitude of change in on/off ration of switching devices was very small. Sayid Bukhari et al. (2020) suggested that oxygen diffusion may play a crucial role due to flow rate and reported at what flow rate they started to observe switching behavior and how it was changing. In our work we have investigated the influence of different $O_2$ flow rate in the narrow rage that supposed to yield best switching behavior, however, we did not observe significant improvement or degradation of film properties. Fig. 4 demonstrates variation of oxygen flow rate between 5 and 25 sccm, all other growth parameters were equivalent for these three curves.

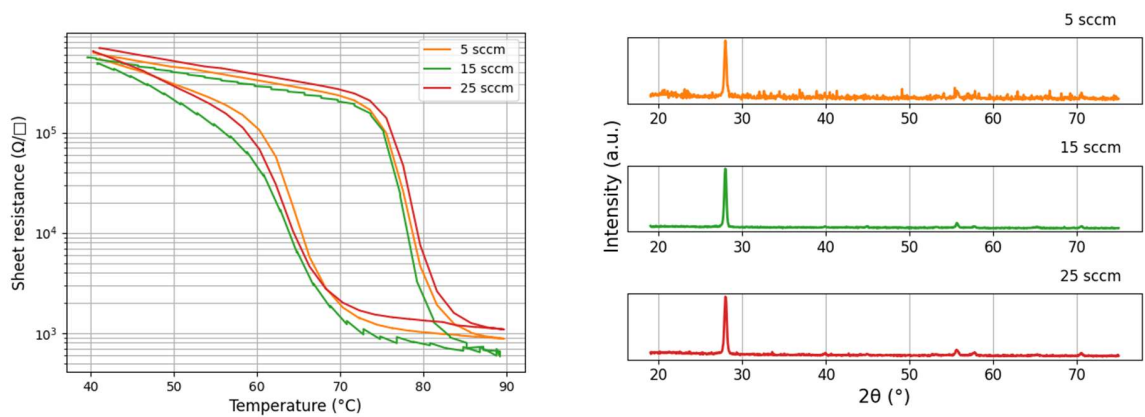

Fig.4. Left: Temperature sweep measured with pour-point prob configuration on samples grown at 5, 15 and 25 sccm of oxygen. Right: respective XRD spectrum for samples grown at 5, 15 and 25 sccm of oxygen. Other growth parameters: T=500C, P=7.5 mTorr, laser frequency=28Hz, laser energy 230mJ, laser fluence=1.01mJ/cm2.

In contrast, a significant influence of total chamber pressure on the quality of $VO_2$ films was observed. Oxygen pressure during both growth and annealing played a critical role in film formation, with a narrow optimal pressure range between 5 and 10 mTorr for PLD growth. Additionally, the total pressure affected the purity of the $VO_2$ phase, as indicated by XRD spectra. A pressure of 6.6 mTorr resulted in the purest $VO_2$ films, correlating with the highest switching ratio compared to other samples, as shown in Fig. 5.

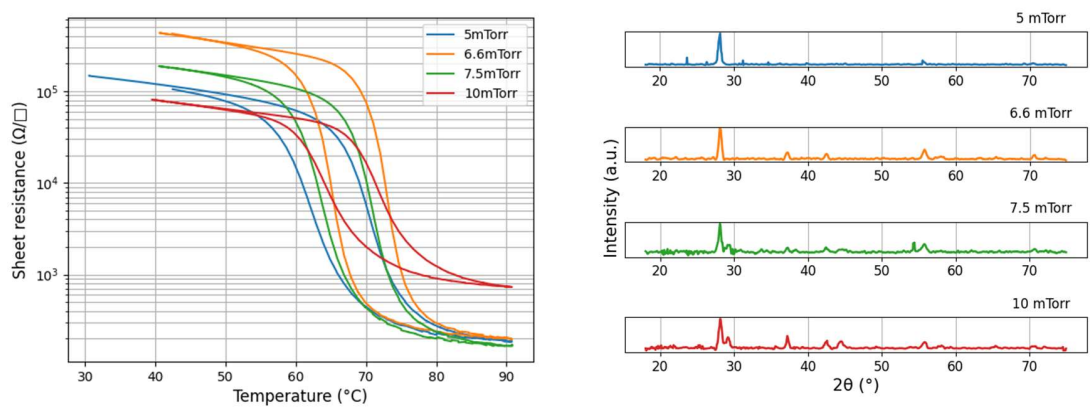

Fig.5. Left: Temperature sweep measured with pour-point prob configuration on samples grown at 5, 6.6, 7.5 and 10 mTorr total chamber pressure. Right: respective XRD spectrum for samples grown at 5, 6.6, 7.5 and 10 mTorr. Other growth parameters: T=400C, laser frequency=28Hz, laser energy 230mJ, laser fluence=1.01mJ/cm2.

LASER ENERGY AND FREQUENCY

Laser energy and frequency were varied to evaluate their impact on the switching properties of the films. Tab. 1. has a list of samples evaluated in terms of frequency difference while energy was kept constant (S1, S2, S3) and samples evaluated in terms of energy while frequency was kept constant (S2, S4, S5, S6) Fig. 5. has thermal switching curves for both sets of samples.

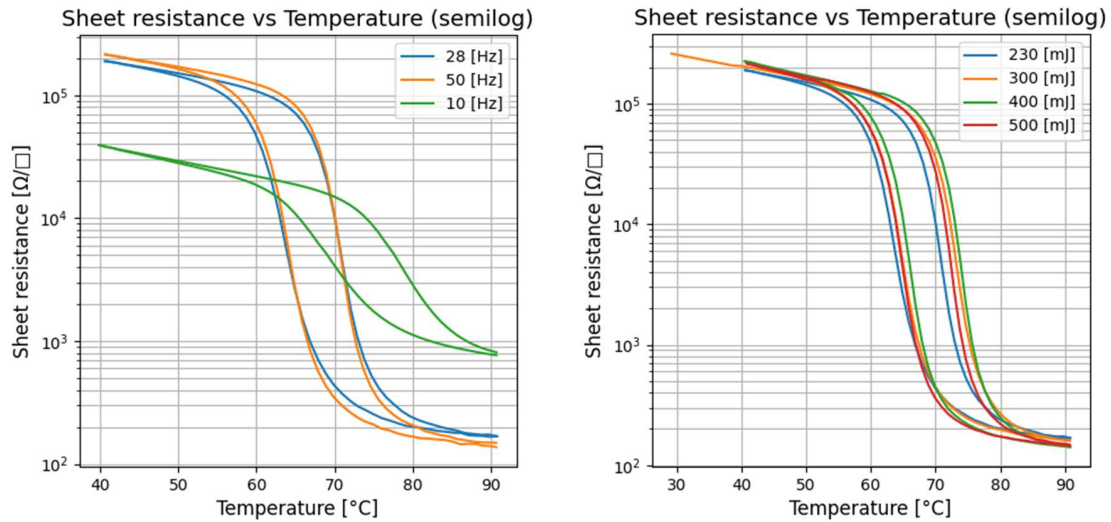

Fig.6. Left: Temperature sweep measured with pour-point prob configuration on samples grown at 10, 28 and 50 Hz. Right: Temperature sweep measured with pour-point prob configuration on samples grown at 230, 300, 400 and 500 mJ. For other growth parameters see table 1.

Even though there are no major difference in switching behavior for samples grown at different energies we undertook further investigation with Raman spectroscopy to control $VO_2$ phase. Raman spectroscopy was used to compare films grown at 10, 28 and 50 Hz as well as films grown at 230, 300, 400 and 500 [mJ] of laser power. Our results for samples with different frequencies of 10Hz, 50zH, and 28Hz show M1 monoclinic insulating phase with no appearance of M2 phase. However, M2 phase appeared with increased energy of the laser. Moreover, coexistence of M1-M2 phase of $VO_2$ seem to result in a higher insulating to metal transition temperature (IMT) at around 73°C.

| Sample #N | T [°C] | f [Hz] | $O_2$ [sccm] | Time [s] | E [mJ] | E [mJ/cm] | [mbar] | Z [nm] | On/Off | IMT [°C] | Peak V-O | Peak V-V |
|---|---|---|---|---|---|---|---|---|---|---|---|---|
| #S1 | 400 | 50 | 15 | 240 | 230 | 1.01 | 0.01 | 81 | 504 | 70.8 | 619.3 M1 | 195, 222 M1 |
| #S2 | 400 | 28 | 15 | 240 | 230 | 1.01 | 0.01 | 69 | 351 | 70.8 | 621.8 M1 | 195.7, 223.7 M1 |
| #S3 | 400 | 10 | 15 | 240 | 230 | 0.99 | 0.01 | 34 | 15.8 | 78.5 | 622.3 M1 | 195.8, 223.2 M1 |
| #S4 | 400 | 28 | 15 | 240 | 400 | 1.77 | 0.01 | 43 | 401 | 73.2 | 619.6, 648.5 M1/M2 | 195.7, 223.7 M1 |
| #S5 | 400 | 28 | 15 | 240 | 300 | 1.31 | 0.01 | 30 | 472 | 74.1 | 619.8, 654.7 M1/M2 | 194.97, 223.2 M1 |
| #S6 | 400 | 28 | 15 | 240 | 500 | 2.23 | 0.01 | 40 | 477 | 72.4 | 619.9, 660.4 M1/M2 | 194.99, 223.13 M1 |

Table.1. List of samples investigated for influence of frequency and energy with complete description of used growth parameters and resulting thickness, on/off ratio, insulating to metal transition temperatures as well as results of Raman spectroscopy with reported V-O and V-V bond peaks corresponding to M1 and/or M2 phase of $VO_2$.

INFLUENCE OF FILM THICKNESS (towards ultra-thin film):

Based on the experimental results, it can be concluded that the optimal conditions for high-quality VO$_2$ thin film growth using the given PLD configuration are: 230 mJ laser energy, 28 Hz frequency, 10 sccm oxygen flow, a total chamber pressure of 6.6 mTorr, and a substrate temperature of 400°C. Additional improvement in the on/off ratio was observed after a 10-minute annealing process at 500°C. This set of parameters will be referred to as the optimized recipe. Subsequent efforts focused on the growth of ultra-thin films and the analysis of the dependence of the on/off ratio and the metal-insulator transition (MIT) on film thickness.

It must be pointed out that at the level of nanometer changes in film thickness and few degrees changes in IMT we must be sure about reproducibility of our results and error bars should be established to evaluate significance of observed dependencies. Therefore, we have repeatedly grown multiple wafers under exact same optimized recipe and calculated that our statistics for on/off ration for optimized recipe are med=473, mean=470, std=76.1 and for optimized plus 10min annealed at 500°C are med=802, mean=835, std=160. We have made same calculations for IMT of optimized recipe and optimized plus 10min annealed at 500°C with med=72°C, mean=71.8°C, std=0.449°C and med=71.8°C, mean=71.8°C, std=0.388°C respectably.

Our optimized recipe has proven to be reproducible and stable and therefore we have investigated possibility of scaling down this film thickness by gradually decreasing time of deposition down to 30 seconds that is minimum allowed time by our PLD machine. Fig.7 demonstrate observed dependency of on/off ratio and IMT temperature for series of sample grown for different durations and therefore thicknesses according to optimized recipe.

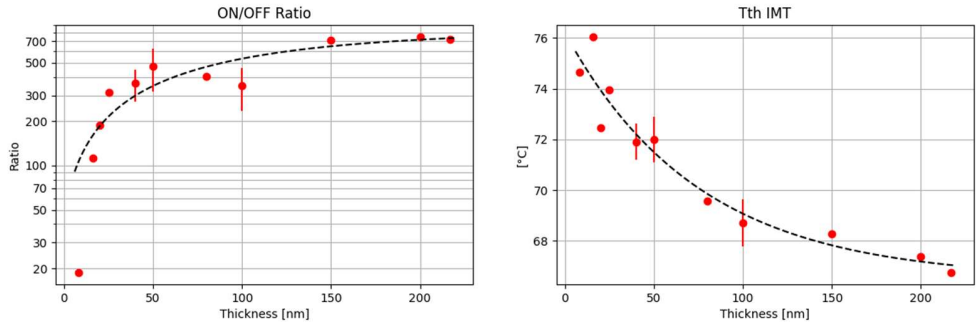

Fig.7. Left: ON/Off ration dependence on sample thickness with exponential fit. Right: Threshold temperature for insulating to metallic transition dependence on sample thickness with exponential fit.

ATOMIC LAYER DEPOSITION METHOD

Atomic Layer Deposition (ALD) is widely regarded as one of the most precise methods for controlling the thickness of thin films, offering sub-nanometer accuracy [37,38]. Its unique, self-limiting surface reaction mechanism enables layer-by-layer growth, ensuring exceptional uniformity across large areas, even on complex or nanostructured surfaces. This capability is particularly valuable for applications requiring highly conformal coatings, such as in advanced microelectronics and 3D architectures, where uniform coverage on high aspect ratio structures is critical. In the context of $VO_2$ thin film fabrication, ALD provides several distinct advantages. It allows for the deposition of ultra-thin, continuous films with excellent uniformity at the wafer scale, which is essential for achieving consistent electrical and thermal properties across devices. Additionally, the process's compatibility with CMOS technology makes it highly attractive for integrating $VO_2$ into silicon-based platforms, facilitating the development of innovative devices, such as phase-change memory, tunable optical components, and neuromorphic circuits.

For this study, all experiments were conducted using a BENEQ TFS200 ALD system, which offers precise control over growth parameters, ensuring reproducibility and scalability. The ability to finely tune parameters such as precursor pulse time, purge cycles, and substrate temperature allows for optimization of film properties, including density, crystallinity, and surface roughness. These characteristics are critical for enhancing the performance of $VO_2$ in switching applications, where uniformity and film integrity directly impact device reliability and efficiency. ALD's capability to produce conformal, ultra-thin layers with high uniformity and its inherent CMOS compatibility underscore its potential as a key enabling technology for next-generation electronic and photonic devices.

ALD depositions reported in this work were run on Tetrakis ethylmethyl amino vanadium (TEMAV) precursor[39] that was heated up to 70°C at bubbler level, water was used as oxygen source, chamber temperature during deposition was kept at 150°C apart from one sample grown at 200°C . Chamber pressure control is not available on our ALD machine and is always set at 4mbar. Thickness of the film was varied by the number of ALD pulses, while pulses were always cycled with 2s of TEMAV and water and with 5s purge in between. At such growth conditions ALD films are always amorphous as deposited and need annealing under correct conditions with temperature and pressure having most significant influence on polycrystalline film formation. Therefore, ALD optimization strategy is mostly based on finding correct pressure and temperature as well as time for annealing step. While temperature range for annealing appears similar to that used in PLD process, pressure range for ALD process is much wider and was evaluated starting from 6.6mTorr and finishing at 75mTorr. Importantly, time of annealing has played a very significant role in film formation. We have experimentally established that lower temperatures and longer times were absolutely critical in formation of ultra-thin continuous ALD films (sample A1 and A2). While higher temperatures and shorter times were acceptable for thicker ALD films (samples A3-A5). Our ALD process optimization was initially based on our previous understanding of film formation during PLD growth, however, more work is needed to develop optimized recipe for ALD. Nevertheless, we have successfully grown and annealed multiple ALD samples with highest applied pressure of 75mTorr resulting at the highest on/off switching ratio. Upper limit of pressure for ALD samples annealing is not reached in our current results and can be further investigated. Fig. 8 shows all 5 successfully grown and annealed ALD samples with corresponding thermal switching curves and XRD measurement showing presence of pure $VO_2$ phase. Sample A1 could not have been measured with our XRD system due to limited sensitivity of detector, however, we can be sure about composition of A1 due to its switching behavior and similarity of growth recipe to other samples. Table 2. summarizes all input parameters for ALD samples.

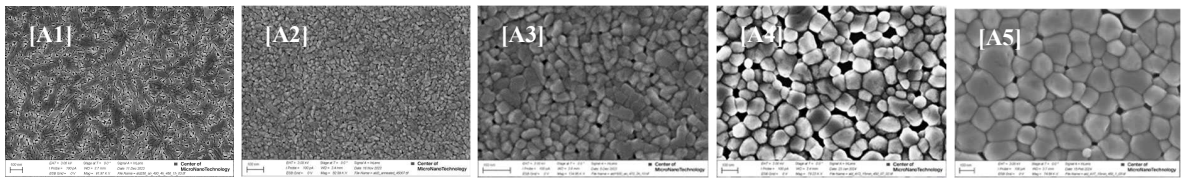

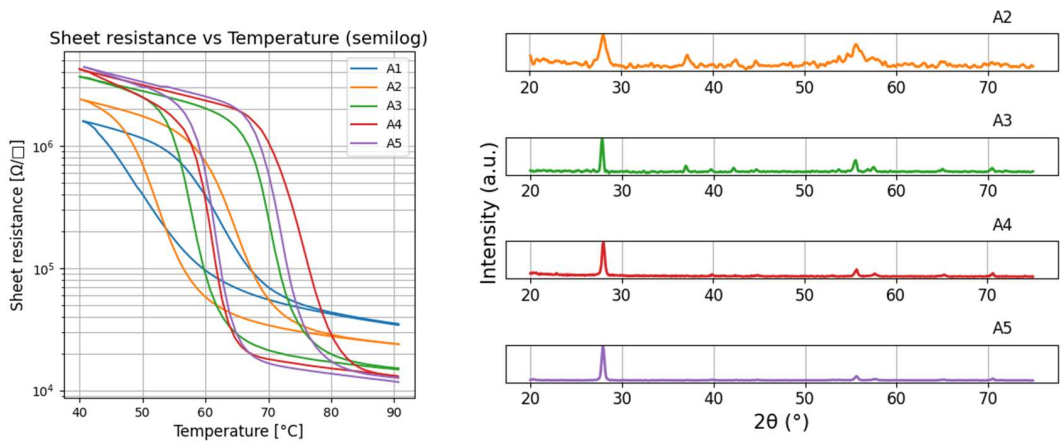

Fig.8. SEM micrographs A1-A5 from left to right and respective thermal switching and XRD curves for A1-A5 ALD samples. Scale bar for all SEM 100nm.

| #N | TEMAV [°C] | Chmb [°C] | # Puls | Ta [°C] | Time Ta [min] | P [mbar] | Z [nm] | IMT[°C] | On/Off |
|---|---|---|---|---|---|---|---|---|---|
| #A1 | 70 | 150 | 250 | 400/450 | 210/30 | 0.0087 | 6 | 62.2 | 16.7 |
| #A2 | 70 | 150 | 790 | 400/450 | 60/120 | 0.0087 | 18 | 64.7 | 35 |
| #A3 | 70 | 150 | 1500 | 470 | 90 | 0.0087 | 30 | 70.3 | 78.6 |
| #A4 | 70 | 150 | 1500 | 450 | 15 | 0.07 | 35 | 75.7 | 106 |
| #A5 | 70 | 200 | 1500 | 450 | 15 | 0.1 | 40 | 72.2 | 136 |

Table.2. List of ALD samples of different thicknesses with growth parameters and resulting thicknesses, on/off and IMT temperatures.

Fig. 9 demonstrate experimentally observed dependency between $VO_2$ film thickness and on/off ratio as well as threshold IMT. These dependencies are very similar to those we have reported for PLD optimized recipe set of films with different thickness.

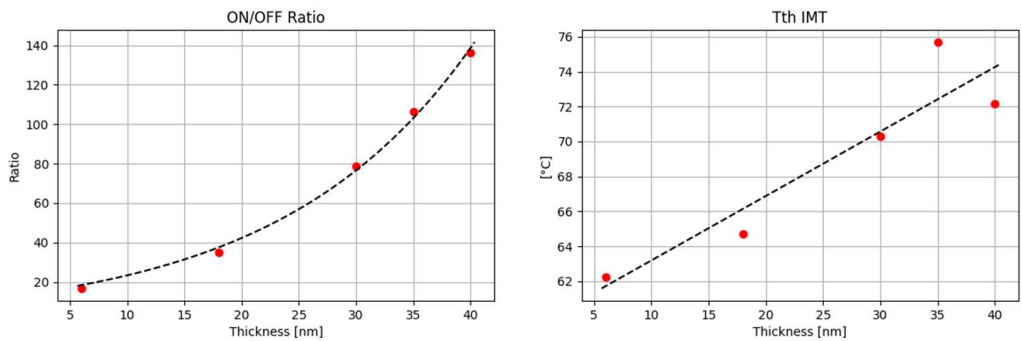

Fig.9. Left: ON/Off ration dependance on sample thickness with exponential fit. Right: Threshold temperature for insulating to metallic transition dependance on sample thickness with linear fit.

Despite need of further investigation into optimized ALD recipe, one of the most interesting results achieved in this work is, indeed, our successful growth of ultra-thin switching polycrystalline $VO_2$ film on CMOS compatible Si wafer with 200nm wetox SiO2. Bellow (Fig.10) we compare ultra-thin 8nm PLD film grown with optimized recipe and ultra-thin 6nm ALD film that could be further optimized. Switching ration for both ultra-thin films are very close to each other and are big enough to be used in $VO_2$ based devices[25,26]. On a contrary, threshold IMT shows significant shift that could be due to difference in grain formation during PLD processes were film is polycrystalline from the beginning and ALD process were films are completely amorphous at growth and turn into polycrystalline only during annealing step. Fig. 10 has also a comparison of optimized 40nm PLD film and best grown so far ALD film. It can be noted that PLD has superior on/off ratio and smaller hysteresis that can serve as a benchmark for further improvements in our ALD recipe.

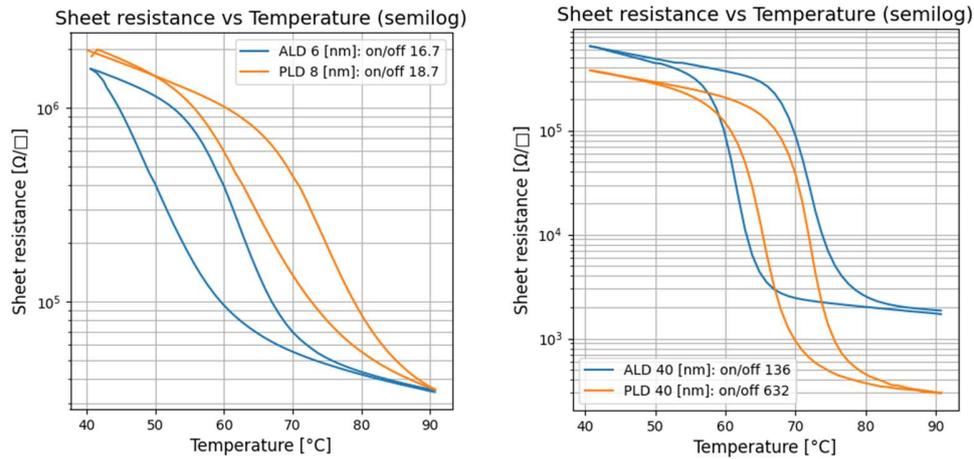

Fig.10. Comparison between thermal switching characteristics of ALD and PLD optimized Left: ultra-thin films and Right: films of 40nm in thickness.

CONCLUSION

A systematic, step-by-step analysis of the influence of growth parameters in pulsed laser deposition (PLD) and atomic layer deposition (ALD) on the formation and properties of CMOS-compatible ultra-thin polycrystalline $VO_2$ films has been conducted. These films, grown on standard $Si/SiO_2$ wafers, were evaluated for electrical properties and surface morphology across a broad thickness range, from 6 nm to 200 nm. Through careful optimization of growth conditions, continuous and functional $VO_2$ films with thicknesses below 10 nm were successfully achieved using both methods. The process demonstrated high reproducibility with low standard deviations across multiple wafers.

The films were characterized using four-point probe measurements in a thermal chamber to assess thermal switching behavior, as well as scanning electron microscopy (SEM), X-ray diffraction (XRD), and Raman spectroscopy for structural and morphological analysis. The metal-insulator transition (MIT) temperature exhibited a thickness-dependent behavior, increasing with reduced thickness in PLD-grown films and decreasing in ALD-grown films. Additionally, the on/off switching ratio decreased with decreasing thickness in both methods.

Temperature and pressure during deposition and annealing were identified as the most critical factors influencing the on/off ratio, MIT threshold, film morphology, and grain size distribution. The ability to achieve CMOS-compatible fabrication of $VO_2$ films highlights the potential for seamless integration into existing silicon-based technologies, enabling the development of low-power, scalable devices for logic, memory, and neuromorphic applications. These findings provide a strong foundation for the design of vertically stacked devices in ferroelectric and memory applications.

AUTHOR INFORMATION

**Corresponding Authors**

*Anna Varini*, Adrian M. Ionescu

Nanoelectronic Devices Laboratory, EPFL, 1015, Lausanne, Switzerland,

anna.varini@epfl.ch, adrian.ionescu@epfl.ch

**Present Addresses**

†Nanoelectronic Devices Laboratory, EPFL, 1015, Lausanne, Switzerland,

**Author Contributions**

Anna Varini, Adrian M. Ionescu, and Igor Stolitchnov prepared the manuscript, interpreted the results, and played a central role in developing the overall scientific methodology for the study.

Anna Varini, Cyrille Masserey, Vanessa Conti, Zahra Saadat Somaehsofla, and Ehsan Ansari carried out all the PLD and ALD depositions, conducted the experiments reported in this work, and contributed to the optimization and analysis of the investigated samples.

**Funding Sources**

This work is supported in part by ERC-2023-SyG SWIMS grant, Number: 101119062, funding from the European Research Council (ERC).

This research was funded in part by the Swiss National Science Foundation (SNSF) Grants: 200021_208233 and CRSII5_209454 Sinergia, Neuromimetic metal-oxide memristors (NeMO).

ACKNOWLEDGMENT

The authors acknowledge the support of CMI-EPFL staff.

ABBREVIATIONS

PLD, Pulsed laser deposition; ALD, Atomic layer deposition; SEM, scanning electron microscopy; XRD, X-ray diffraction; MIT metal-insulator transition; TEMAV, Tetrakis ethylmethyl amino vanadium.